# Diagnostics of low pressure hydrogen discharge created in a 13.56 MHz RF plasma reactor

**J. Krištof**[1], **A. Annušová**[1], **M. Anguš**[1], **P. Veis**[1], **X. Yang**[2], **T. Angot**[2], **P. Roubin**[2] **and G. Cartry**[2]

[1] Department of Experimental Physics, Faculty of Mathematic, Physics and Informatics, Comenius University in Bratislava, Mlynská dolina F2, 84248 Bratislava, Slovak Republic

[2] Aix-Marseille University, CNRS, PIIM, UMR 7345, 13013 Marseille, France

E-mail: jaroslav.kristof@gmail.com, adriana.annusova@fmph.uniba.sk

**Abstract.** A 13.56 MHz RF discharge in hydrogen was studied within the pressure range of 1-10 Pa, and at power range of 400 - 1000 W. The electron energy distribution function and electron density were measured by a Langmuir probe. The gas temperature was determined by the Fulcher-α system in pure $H_2$, and by the second positive system of nitrogen using $N_2$ as the probing gas. The gas temperature was constant and equal to 450 ± 50 K in the Capacitively Coupled Plasma mode (CCP), and it was increasing with pressure and power in the Inductively Coupled Plasma mode (ICP). Also the vibrational temperature of ground state of hydrogen molecules was determined to be around 3100 and 2000 ± 500 K in ICP and CCP mode, respectively. The concentration of atomic hydrogen was determined by means of actinometry, either by using Ar (5 %) as the probing gas, or by using $H_2$ as the actinometer in pure hydrogen (Q1 rotational line of Fulcher-α system) The concentration of hydrogen density was increasing with pressure in both modes, but with a dissociation degree slightly higher in the ICP mode (a factor 2).

## 1. Introduction

In tokamaks (nuclear fusion reactors), a hot plasma composed of deuterium and tritium nuclei is magnetically confined to achieve fusion to produce energy. The ITER tokamak project aims to demonstrate the feasibility of producing energy, while the DEMO reactor aims to be the first fusion reactor prototype to deliver power to the electric network. The present project is put into the context of these international projects. In tokamaks the core plasma is hot, about 1 – 10 keV, and the edge plasma is colder and has a temperature of few eV, depending on the tokamak. The edge plasma, despite being colder, strongly interacts with the walls of the machine. Particular parts of the walls, e.g. divertors, are subjected to high heat and particle fluxes leading to material erosion. Considering the case of erosion, divertors made of low atomic number (low-Z) material have, until now, been preferred due to the limit of energy loss by radiation. During recent years, graphite has been studied extensively, as material of which divertors would be composed, due to its invaluable thermal and mechanical properties (Mellet *et al* 2014, Pardanaud *et al* 2014, Pardanaud *et al* 2013, Pardanaud *et al* 2012, Bernard *et al* 2013, Martin *et al* 2011, Tsitrone *et al* 2011, Pardanaud *et al* 2011, Richou *et al* 2007). However, severe drawbacks in the form of high retention of tritium (Pegourie *et al* 2013, Roth *et al* 2009) and low resistance to neutron bombardment as well as erosion and co-deposition of hydrocarbons in shadowed area mean that graphite will never be used as a divertor material in fusion reactors. Therefore, current research focus is put on high Z materials, with tungsten being considered as the most appropriate candidate. As it stands, both the German tokamak ASDEX upgrade (Hermann and Gruber 2003) and the European JET (EUROfusion 2014-2018), currently the most powerful tokamak in the world, are now using tungsten-coated divertors. Also the project WEST (Grosman *et al* 2013, Marandet *et al* 2014), aims to study and build a tungsten divertor in the tokamak Tore Supra at Cadarache. High-Z materials possess the advantage of having a high sputter threshold in hydrogen plasma, but at the cost of radiating a huge power and extinguishing the plasma in case of impurity events. To prevent tungsten erosion in fusion reactor, it is crucial to reduce the heat load on divertors to less than 10 MWm$^{-2}$ in steady state. Impurity gas seeding (N$_2$, Ar, Ne…) is considered as the primary technique to decrease the heat load to the divertor.

In fusion device, plasma facing components play a major role in recycling species. The edge plasma physics and the plasma-wall interactions are important bottlenecks for the development of fusion energy. Current edge codes used for the ITER design process, such as SOLEDGE2D-EIRENE (based on a fluid approach for the plasma and on kinetic Monte Carlo simulations for neutral particles) (Ciraolo *et al* 2014), have only very limited ability to deal with mixed material environments (Toma *et al* 2010, Bonnin *et al* 2009), due to the lack of comprehensive wall model. These codes can to some extent describe impurity migration in the machine, in order to simulate the change of the wall composition. But still, reliable data are not yet available for: thermal properties, sputtering yields and reflection coefficients for the resulting mixed materials. Among the data needed for simulation codes, atomic H reflection by a surface under plasma exposure is a key parameter playing a strong role on the process of hydrogen recycling at plasma wall (Brezinsek *et al* 2005, 2003). Under the detached plasma conditions that will be encountered at ITER divertor, ion energy is below 20 eV. The TRIM code (Eckstein 1991) which was used in the past to calculate atomic H reflection coefficients, is no longer useful since it is based on the binary collision approximation which ceases to be valid below few tens of eV. More recent calculations use molecular dynamic (MD) simulations (Henriksson *et al* 2006, Inagaki *et al* 2011) instead of TRIM calculations to determine the reflection probability at low energy. As MD simulations rely on a calculated interaction potential, results obtained must always be benchmarked with experiments. Finally, none of these calculations take into account impurity seeding. Therefore we have developed a dedicated laboratory plasma experiment to measure hydrogen (or deuterium) reflection coefficient and its dependence on various parameters (impurities, wall temperature, ion flux, atomic flux...).

The present paper deals with the characterization of the above mentioned hydrogen plasma source, which is required before any measurement of the reflection coefficient. Let us note that this thematic of atomic loss (or atomic reflection) on surface is not only limited to fusion related studies but concerns most of the low-pressure plasma processes. The measurement of the reflection coefficient and its dependence with experimental parameters is a recurrent issue in many plasma applications ranging from microelectronics to fusion, as evidenced by the high number of papers dealing with this subject over tens of years (Wood and Wise 1962, Cartry *et al* 2000, Rousseau *et al* 2001, Macko *et al* 2004, Lopaev and Smirnov 2004, Mozetic and Cvelbar 2007, Kurunczi *et al* 2005, Rutigliano and Cacciatore 2011, Jacq *et al* 2013, Bousquet *et al* 2007, Guerra 2007, Kang *et al* 2011, Samuell and Corr 2014, Sode *et al* 2014, Marinov *et al* 2014 a). Finally, hydrogen plasma characterization is of interest not only because of the atomic surface loss issue, but also because hydrogen gas is often used for plasma processing such as hydrogenation (Hatano and Watanabe 2002), treatment of Si wafers for creating subsurface defects in layers (Ghica et al 2010), chemical vapour deposition of diamonds (Hassouni *et al* 1996), functional materials or polycrystalisation of amorphous Si, creation of positive (Liu and Fonash 1993, Cielaszyk *et al* 1995) or negative-ions in ion-sources (Iordanova *et al* 2011, Kalache *et al* 2004, Ahmad *et al* 2010).

In this work we characterize the RF plasma source in Capacitively Coupled (CCP) and Inductively Coupled (ICP) Plasma modes by studying primarily a pure hydrogen plasma. We determine gas temperature using optical emission spectroscopy under nitrogen admixture, electron density and temperature using Langmuir probe, vibrational temperature of ground state of hydrogen by using Fulcher-α emission system, and atomic hydrogen concentration using actinometry under argon admixture. All measurements were realized versus pressure (1 - 10 Pa) and delivered power (400 - 1000 W) in continuous discharge or pulsed mode. The experimental apparatus is first described, and followed by the description and discussion in particular subsections of the methods and the experimental results obtained.

## 2. Experimental set-up and conditions of measurements

The scheme of the experimental set-up is given in figure 1. The plasma source consists of a spherical stainless chamber 300 mm in diameter. A planar coil antenna (three turn) is separated from the plasma by a quartz window (166 mm in diameter, 12 mm thick) placed on top of the plasma chamber. The antenna is connected to an L-type matching box connected to Advanced Energy Cesar 1310 RF generator. The plasma chamber is made of stainless steel with Pyrex and Quartz windows and molybdenum sample holder. The gas is introduced in the plasma chamber through mass flow controllers and pumped thanks to a 550 l/s turbomolecular pump followed by an oil rotary pump. The base pressure in the plasma chamber is $10^{-6}$ Pa (measured by ionization gauge). We studied pure hydrogen plasma at powers 400, 600, 850 and 1000 W and at pressures from 1 to 10 Pa. The pressure during the plasma is measured by a capacitive gauge and is regulated by changing the pumping speed by modifying a gate valve opening placed between the plasma and the turbomolecular pump.

A sample holder from Vegatec company is used to hold 4'' wafers. The sample holder allows for wafer temperature measurement thanks to a backside thermocouple, wafer heating and cooling (from room temperature to 800°C) using backside resistive heater and backside circulation of fluid (air or glycol-based liquid), wafer loading and clamping (the chamber is equipped with a load lock chamber), and biasing of the wafer down to -1 kV. The sample holder can be translated vertically. It was adjusted in a way so the surface of the sample would be at the position 0 cm on figure 1. In order to limit the plasma expansion and better define the wall conditions for surface loss measurements, a moving Pyrex tube (external diameter 170 mm, thickness 5 mm, height 146 mm) has been installed in the plasma chamber. When the tube is in the upper position, the plasma is confined by the tube itself, the quartz window on top and the sample holder on bottom. Pumping openings (~10 mm typical dimension) remain on top between the tube and the quartz window, and on bottom between the sample holder and the tube. At this position the middle of the tube is 5 cm from the bottom side of the reactors quartz window (also from the sample holders surface). When the tube is in the lower position, the plasma created between the top antenna and the wafer holder extends towards the stainless steel walls. Most of the characterizations performed in this paper have been performed without the tube. For the purpose of spatial characterisation of the plasma, a 4'' $SiO_2$ was loaded on the sample holder and heated to 450 K during measurements. $SiO_2$ was chosen as a reference material because of a large number of paper dedicated to atomic hydrogen loss on this material (Cartry *et al* 2000, Bousquet *et al* 2007, Kim and Boudart 1991, Kae-Nune *et al* 1996).

In all cases, the gas flow rate is fixed at 10 sccm. For atomic hydrogen determination by actinometry an admixture of 5 % of argon was used in hydrogen. For the determination of the gas temperature 5, 10, 15, 30, 50 % of admixtures of nitrogen were used in hydrogen. Signal is collected by an optical fibre through quartz lens from the chamber. The collected signal is recorded with an Andor Mechelle ME5000 spectrometer coupled with an Andor iStar intensified camera (DH734) in wavelength ranging from 215 to 950 nm. Experimental spectra were corrected by the spectral response of spectrometer. Resolution of Mechelle spectrometer is $\lambda/\Delta\lambda = 4000$.

Plasma parameters as electron density and electron temperatures were measured by RF-compensated SmartSoft Langmuir probe with a tungsten tip of 10 mm long and 200 μm in diameter.

Langmuir probe was translatable and was positioned at the centre of the plasma chamber at a height of 5 cm from the bottom part of the quartz window.

The transition from CCP to ICP modes is identified and marked for all evolutions shown here. Note that this transition, characterized by a hysteresis effect, occurred with a sudden change of both the emission intensity and the plasma density and it was triggered by changing the delivered power.

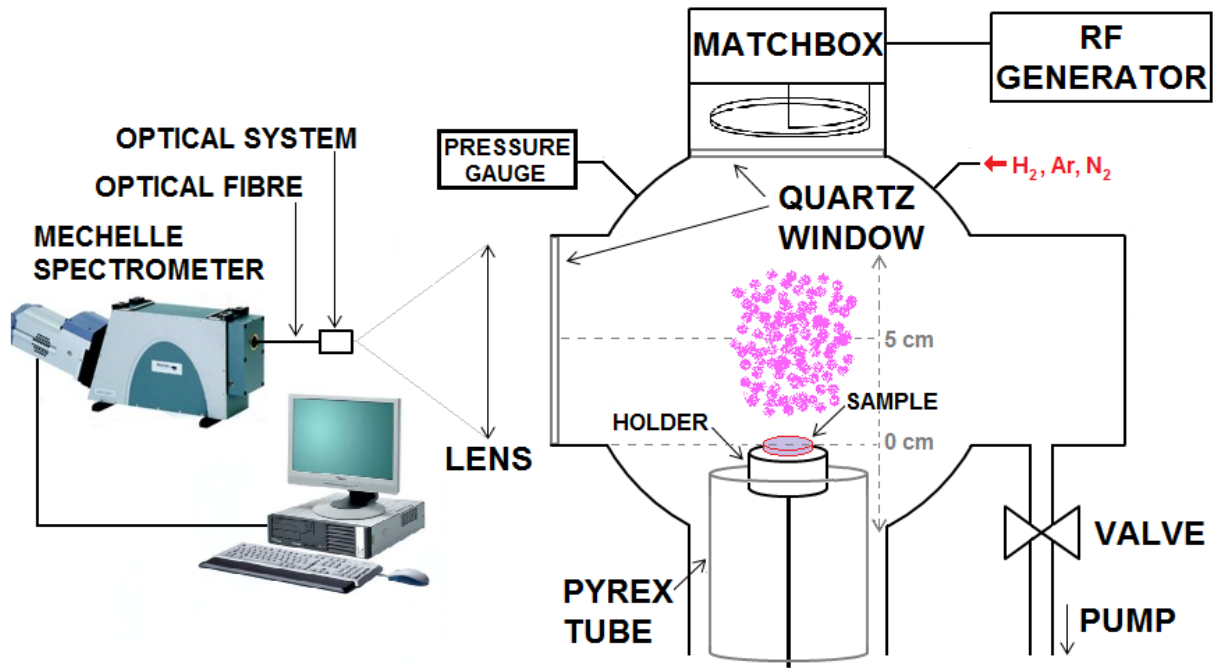

**Figure 1.** Illustration of the experimental set-up.

## 3. Results and discussion

*3.1. Electron density and electron temperature*

Measured data with the Langmuir probe (placed in the centre of the chamber/plasma, height: 5 cm from the windows lower part) were processed by software SmartSoftV5.01 from Scientific systems (Hopkins 1995). The electron temperature and electron density were determined with the help of the plasma potential, obtained using the zero second derivatives. The figure 2 is depicting the measured electron densities in pure $H_2$ as function of power and pressure. Overall, the electron densities are increasing by increasing power in the measured range of pressure, i.e. between 1 and 10 Pa. The transition to the ICP mode is characterized by a sudden jump in the electron densities above ~ $10^{10}$ cm$^{-3}$. With increasing pressure this transition occurs at lower values of power. The mentioned hysteresis effect (Daltrini *et al* 2008, Kang and Gaboriau 2011) is evidenced, as for example at 4 Pa and 600 W the discharge can be both in ICP and CCP mode.

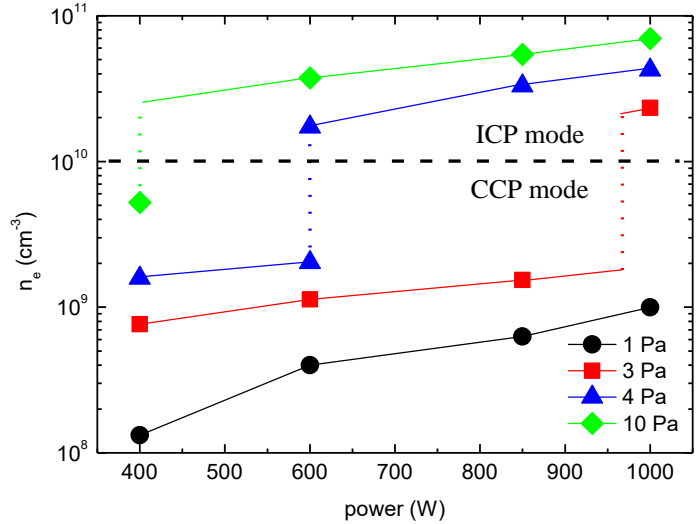

**Figure 2.** Electron density in pure $H_2$ plasma as function of pressure for powers between 400 and 1000 W. Overall, the electron densities are increasing by increasing power. The transition from CCP to ICP discharge mode is marked with dashed line. With increasing pressure this transition occurs at lower values of power.

The figure 3 presents the electron concentrations in the case of a $H_2$-$N_2$ plasma at the two pressure limit conditions, at 1 (a) and 10 Pa (b). We are reminding that the admixture of nitrogen was realized in order to help determining the gas temperature. However this affected the discharge properties and led to some changes in the electron concentration compared to the case of pure hydrogen. First of all the $H_2$-$N_2$ plasma is also characterized by rising values of $n_e$ with both pressure and power as in pure hydrogen and by a sudden increase of $n_e$ concerning the transition from CCP to ICP mode. However at 1 Pa nitrogen causes important increase of the electron concentrations and even transition to the ICP mode at 1000 W and 50 % $N_2$ unlike in the case of pure $H_2$. Increasing of electron density with nitrogen density was observed before and explained by higher ionization cross section of $N_2$ in comparison with $H_2$ (El-Brulsy *et al* 2012). In contrast, at 10 Pa approximately the same trend is observed for all dependences as in pure hydrogen, i.e. no important changes are induced with nitrogen addition.

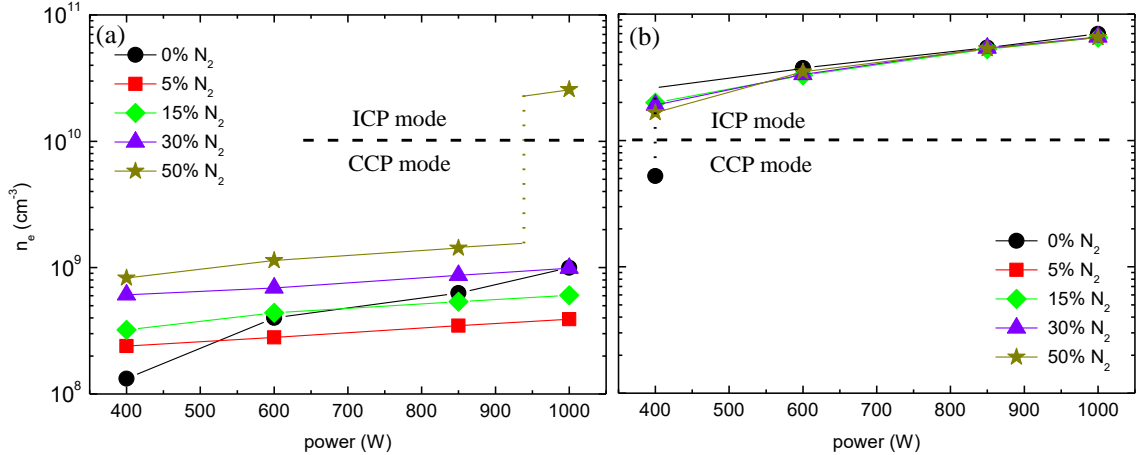

**Figure 3.** Electron density in mixture of $H_2$-$N_2$ at 1 Pa (a) and 10 Pa (b) as function of power for different $N_2$ percentages. The transition from CCP to ICP discharge mode is marked with dashed line. Nitrogen addition was realized to help determine the gas temperature, however this led to some changes in the electron concentration compared to the case of pure hydrogen.

The Electron Energy Distribution Function (EEDF) in pure $H_2$ plasma is depicted in figure 4 for different pressure and power conditions. Note, that there were no important changes with power or mode of plasma for what concerns the shape of the EEDF. Maxwellian EEDF was obtained for lower pressure and it was changed to bi-Maxwellian for pressures higher than 4 Pa. The electron temperatures marked in the graph are merely informative. With decreasing pressure we see an increase of the high electron energy tail to the disadvantage of the lower energy part. In hydrogen the transition with pressure from a Maxwellian to a bi-Maxwellian EEDF has been studied previously in a 13.56 MHz CCP discharge (Abdel-Fattah and Sugai 2013) and a bi-Maxwellian distribution reported at 10 Pa in a pulsed ICP discharge (Osiac *et al* 2007).

Measured EEDF was used for calculation of rate constants used in actinometry (paragraph 3.4) and in calculation of vibrational temperature (paragraph 3.3).

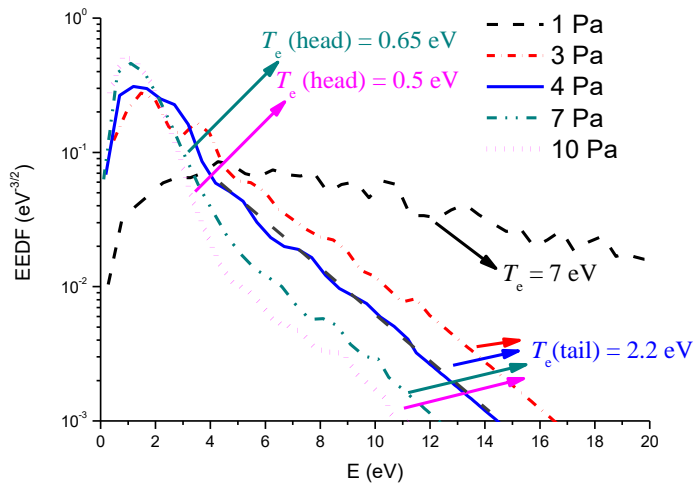

**Figure 4.** EEDF and the deduced electron temperatures as function of pressure at power of 400 W for 1 Pa and 1000 W for the higher pressure conditions. The temperatures are obtained assuming a Maxwellian or bi-Maxwellian distribution.

*3.2. Gas temperature*

As rotational-translational relaxation is non-adiabatic, fast equilibrium can occur between rotational and translational degrees of freedom after few collisions (Britun *et al* 2007). For this reason, rotational temperature of ground states of molecule is very often taken as gas temperature. One of the most known emission system of hydrogen discharge in the visible range is Fulcher-α system ($H_2(d^3\Pi_u^-) \rightarrow a^3\Sigma_g^+$)). For gas temperature determination often the Q-branch of this system is used as other branches, P and R, are perturbed by Σ states (Qing *et al* 1996). Rotational temperature of $H_2(d^3\Pi_u^-)$ state determined from hydrogen Fulcher-α band is not very often in equilibrium with gas temperature at low pressures because of very short lifetime of the $H_2(d^3\Pi_u^-)$ state. So $H_2(d^3\Pi_u^-)$ cannot be thermalized. Furthermore hydrogen needs approximately 300 collisions to achieve equilibrium (Farley *et al* 2011) while 5-10 collisions are required for nitrogen. According to Tomasini *et al* (1996), the collision frequency as a function of the pressure is given by: $\upsilon_{coll}/p = 8 \times 10^6$ s$^{-1}$Torr$^{-1}$. The emission probability of $H_2(d^3\Pi_u^-)$ is $4 \times 10^7$ s$^{-1}$. If $\upsilon_{coll}$ is much lower than the emission probability, as in our experimental conditions, the equilibrium cannot be ensured. However, under our experimental conditions, the excited state $H_2(d^3\Pi_u^-)$ is mainly populated through electron collisions from the ground state. If the electron impact rate coefficient is constant over the rotational levels, and if the transitions without change of angular momentum are dominant, then the rotational distribution of the $H_2(d^3\Pi_u^-)$ is a copy of the ground state distribution, and the rotational temperature can be determined from Flucher-α band. It has been checked that these assumptions can be considered as valid for the Q branch of Fulcher band (De Graaf 1994, Gans *et al* 2001). Therefore, the rotational temperature determined from this band is a good approximation of the rotational temperature of the ground state. Then it is assumed that rotational and translational temperature (gas temperature) are in equilibrium. Intensity of rotational line is given as (*N* quantum number can be used instead of *J*)

$$I \propto (2\Gamma_{N'} + 1) S_{aN''}^{dN'} exp\left(\frac{-F_{dv'N'}hc}{kT_{rot}}\right), \quad (1)$$

where $S_{aN''}^{dN'} = \frac{2N'+1}{2}$ is Hönl-London factor, $T_{rot}$ is rotational temperature of the state and $F_{dv'N'} = B_v N(N+1)$ is the rotational energy. The total nuclear spin due to presence of orto- ($\Gamma_{N'}= 1$) and para-states ($\Gamma_{N'}= 0$) of H$_2$ is taken into account, where $\Gamma_{N'} = 1$ for *N'* even and $\Gamma_{N'} = 0$ for *N'* odd. Rotational temperature of $H_2(d^3\Pi_u^-)$ is then determined assuming a Boltzmann distribution of rotational states. The rotational constant of $H_2(X^1\Sigma_g^+)$ has to be used for the calculation (since the rotational distribution of the excited state is a copy from the ground state one). However, as the rotational constant of $H_2(X^1\Sigma_g^+)$, $B_{vg}$, is two times higher than the rotational constant $B_v$ of $H_2(d^3\Pi_u^-)$ state ($B_{vg} = 2B_v$), then it simply applies: $T_{rot}(H_2(X^1\Sigma_g^+)) = 2\ T_{rot}(H_2(d^3\Pi_u^-))$. For the rotational constant we can write: $B_v = B_e - \alpha_e \left(v + \frac{1}{2}\right)$ with $B_e = 30.364$ cm$^{-1}$, $\alpha_e = 1.545$ cm$^{-1}$ for the $H_2(d^3\Pi_u^-)$ state and $B_{eg} = 60.809$ cm$^{-1}$ $\alpha_{eg} = 2.993$ cm$^{-1}$ for $H_2(X^1\Sigma_g^+)$ state (Herzberg 1955). An example is illustrated on figure 5 for the vibrational transition *v'- v''* = 1-1 of the Q-branch of the Fulcher-α band.

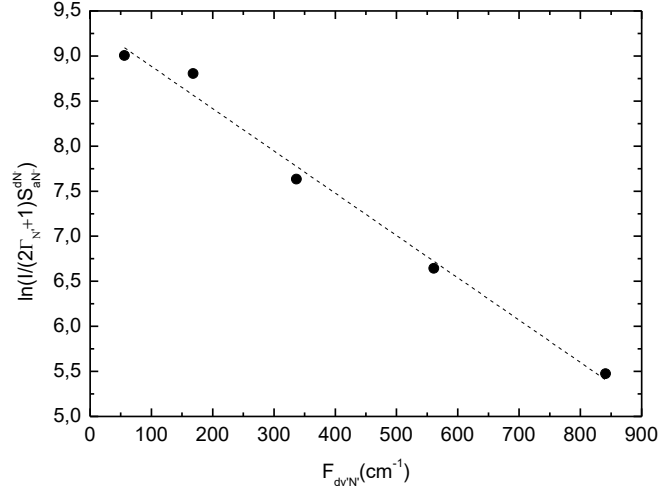

**Figure 5.** Boltzmann plot of rotational states for the vibrational transition $v'$-$v''$ = 1-1 of the Q-branch of the Fulcher-α band (pure $H_2$, 4 Pa, 1000 W).

It was observed by Fantz (2002) and also by Tsankov *et al* (2012) that temperatures determined from different diagonal bands are not equal. This observation is not explained, yet. Fantz compared different spectroscopic techniques and she found out that temperature determined by vibrational bands 1-1 and 2-2 are the most close to the real gas temperature. For this reason we concentrated on these bands. We have chosen the transition 1-1 (at λ = 612-616 nm) because of the higher intensity of rotational lines compared to the vibrational band 2-2. Note, that the experimental error on the determination of the rotational temperatures is ± 50 K, coming from the accuracy of the fit. It means that the error for the gas temperature from the Fulcher-α system is ± 100 K. The determined gas temperatures in pure hydrogen are depicted on figure 6 (a). Gas temperature in CCP mode was equal to an average value of 450 ± 50 K and did not depend on pressure or on delivered power. It is a non-heating mode where a large portion of energy goes accelerating the ions in the plasma sheath and is not used to heat the gas through electron collisions. On the other hand, ICP mode can be characterized by collisional heating. In this so called heating mode the gas temperature was increasing because more energy was transferred to the plasma with increasing power and with pressure, the plasma density was increasing leading to more frequent collisions and larger energy transfer from electrons to neutral gas. Gas temperature was increasing with highest slope at 10 Pa from 525 (400 W) to 800 K (1000 W). Similar values were obtained recently in a high density low-pressure hydrogen discharge (Samuell and Corr 2015).

In the case of nitrogen admixture, we compared the gas temperatures determined by the Fulcher-α system and the second positive system of nitrogen ($N_2(C^3\Pi_u \rightarrow B^3\Pi_g)$), noted 2PS. We suppose that rotational temperature of 2PS is equal to gas temperature (Fantz 2004, Behringer 1991). In the latter case the vibrational transition $v'$-$v''$ = 1-0 (at λ = 315.8 nm) was used being the most intense one in the spectra after the transition 0-0, which however overlapped with the $NH(A^3\Pi - X^3\Sigma)$ system. By spectral simulations using the software SPECAIR (Laux 2002) we determined the rotational temperature of the $N_2(C^3\Pi_u)$ state which can be considered as the gas temperature with an error of ± 50 K. The Fulcher-α system could be used for the determination of $T_g$ only at 5, 10 and 15 % of nitrogen added, as at higher percentages of $N_2$ (30 and 50 %) the first positive system of nitrogen ($N_2(B^3\Pi_g \rightarrow A^3\Sigma_u^+)$) overlapped with the bands from Fulcher-α. Concerning the pressure dependence of the gas temperature in the case of nitrogen admixtures, the behaviour is equivalent to that in pure hydrogen. Therefore the figure 6 (b) presents the results as function of power and nitrogen percentage in the mixture only in the case of the pressure of 10 Pa. First of all, the results using 2PS and the Fulcher-α system are in quite good agreement. Even if the error bar are quite large, it reinforces the confidence in the gas temperature determination from the Fulcher-α system. There is only one noticeable difference between the gas temperatures determined from both systems: at 10 Pa and 1000 W the second positive system is giving a temperature lower than the Fulcher system. The discharge was in ICP mode for powers higher than 400 W under the presence of nitrogen admixture. At 400 W, discharge can be in ICP and CCP mode in pure hydrogen and also with 5 % of nitrogen. The transition causes additional heating of the gas as stated before. Thus for H$_2$-N$_2$ mixtures also higher gas temperature values were determined in the ICP mode. For a given power the gas temperature changes with nitrogen admixture seems to be related to the uncertainty in the measurements rather than to a real change of heating mechanism. On the average, in ICP mode the gas temperature in mixtures is close to the value determined in pure hydrogen discharge. In CCP mode with N$_2$ admixture $T_g$ was equal to the value determined for pure hydrogen within the determination uncertainty (see the case of 400 W).

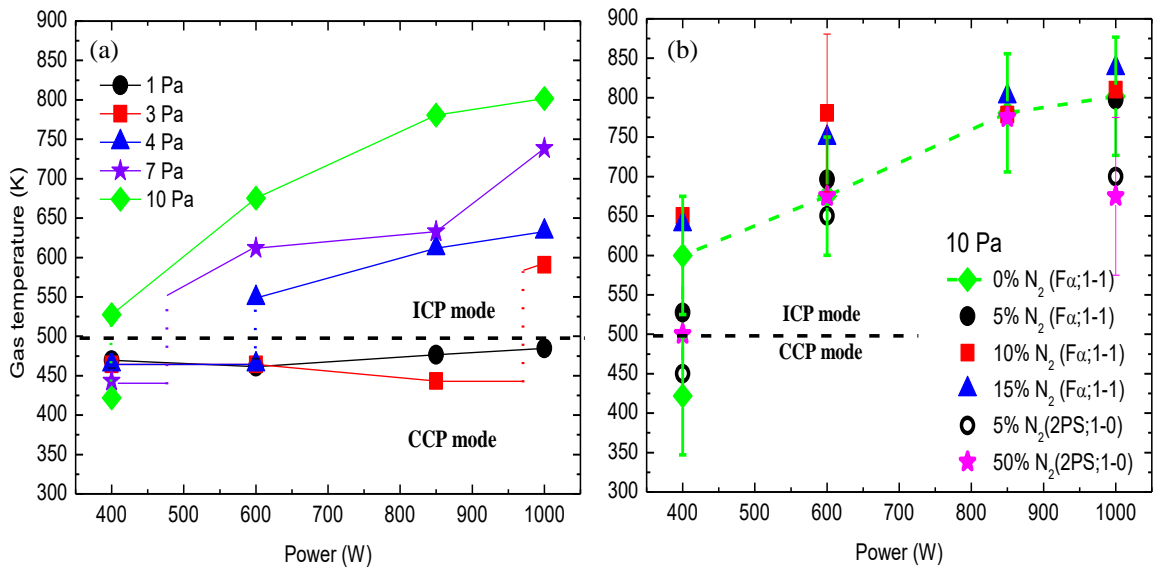

**Figure 6.** Gas temperature in pure hydrogen as function of pressure determined from Boltzmann plot of vibrational band 1-1 of Fulcher-α (a). Comparison of gas temperatures at 10 Pa as function of delivered power in the case of nitrogen admixtures determined from vibrational band 1-1 of Fulcher-α for 0 5, 10 and 15 % N$_2$ (full

circles) and from the band 1-0 of the second positive system of nitrogen for 5 and 50 % N$_2$ (empty circles) (b). The gas temperature is noticeably higher in the inductive mode than in the capacitive mode where no significant heating is measured. Gas temperature variations measured with Fulcher-α and second positive system are in quite good agreement.

For determination of the reflection coefficient of atomic hydrogen, it is possible to use pulsed-discharge and measure the atomic decay in time afterglow. As the gas temperature can influence the decay of atomic hydrogen, we have performed to time resolved measurements of the gas temperature during pulsed plasmas. A duty ratio of ON:OFF = 50 ms: 50 ms was used for investigation of the temporal evolution of gas temperature. The $T_g$ values were determined by 2PS and Fulcher-α systems, as described above and using the transitions mentioned previously. In the pulsed regime, for a gas mixture with 50 % nitrogen the Fulcher-α system was not influenced by the first positive system of nitrogen and could also be used. The results are presented on figure 7 for two measured pressure conditions. The temperature is increasing during the pulse and reaches a stationary value after ~1.5 ms. At that time the value of $T_g$ in the continuous discharge is achieved. Note that the 2PS at 10 Pa and 1000W is giving a temperature noticeably lower that the Fulcher system, as already observed in the continuous mode. It is not possible to determine gas temperature during the OFF phase because of short lifetime of excited states.

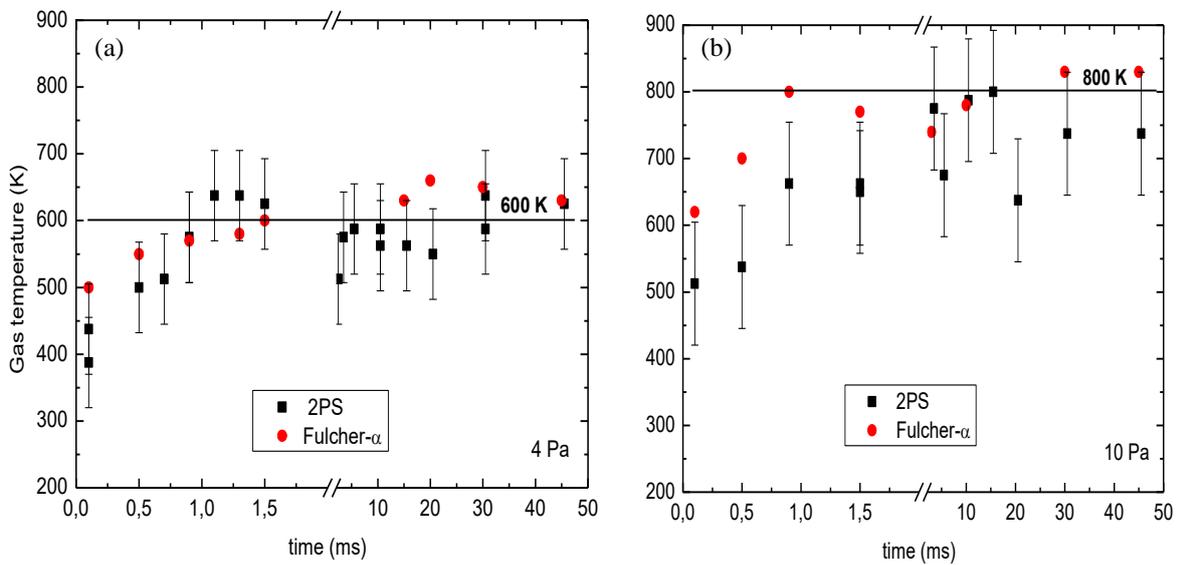

**Figure 7.** Evolution of gas temperature during one pulse (ON:OFF = 50ms:50ms) at 1000 W in 50 % nitrogen admixture at pressures of 4 (a) and 10 Pa (b). Black line shows gas temperature in the case of the continuous discharge. The temperature is increasing during the pulse and reaches a stationary value after ~1.5 ms, which corresponds to the value of $T_g$ in the continuous discharge.

The reflection coefficient can be strongly influenced also by the temperature of the sample which can in turn also change the gas temperature. At low pressures the gas particles collide with the surfaces of the chamber much more often than between each other and their loss depends on the surface temperature. The spatial variation of gas temperature towards the sample was measured both with and without Pyrex tube at 1000 W and pressures of 4 and 10 Pa while the sample was heated to 450 K or maintained at 300 K by liquid cooling. If the temperature of sample is changed, gas temperature will be changed according to the difference of sample temperature. This is in agreement with figures 8 (a) and (b), which present the evolution of the gas temperature at 4 and 10 Pa, respectively, with and without the Pyrex tube and at different sample temperatures. For the determination of $T_g$ the same transition of the Fulcher-α system was used as in the cases above. The change of sample temperature about 150 K changed gas temperature approximately from 650 K to 800 K at 4 Pa (without tube) and from 800 K to 950 K at 10 Pa (without tube). The Pyrex tube delimited the chamber and also the discharge to smaller dimensions, thus the power density was higher. That induced higher gas temperature when Pyrex tube was used. Difference between $T_g$ with and without Pyrex tube is lower at 10 Pa because plasma was not that spatially extended at higher pressure without the tube.

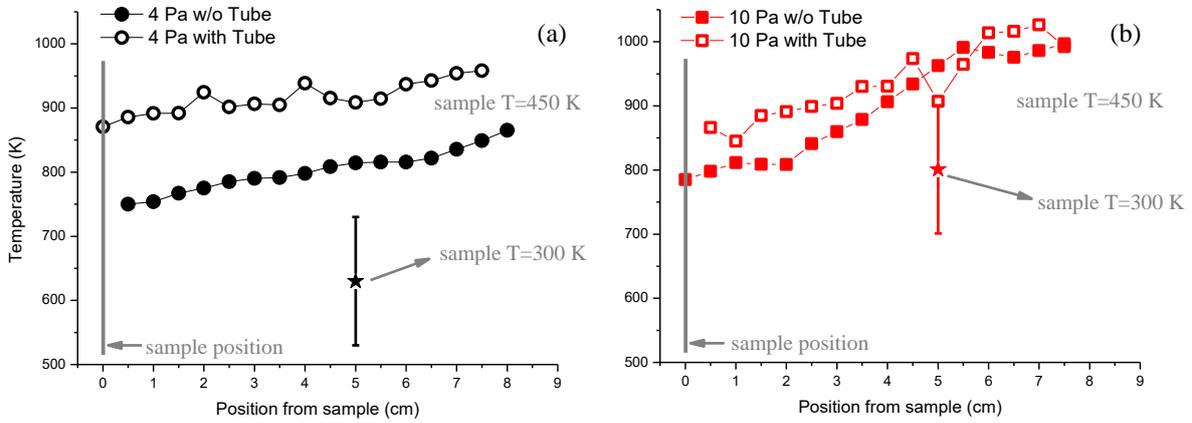

**Figure 8.** Evolution of gas temperature at 4 Pa (a) and 10 Pa (b) with (empty symbols) and without Pyrex tube (full symbols). The points marked by stars are measured without Pyrex tube and heating of sample to 300 K. Other measurements are done at 450 K of sample holder. The Pyrex tube delimited the chamber and also the discharge to smaller dimensions, increasing the power density resulting in higher temperatures of the gas.

*3.3. Vibrational temperature of the ground state $H_2(X^1\Sigma_g^+)$ and relative negative ion densities*

The knowledge about the vibrational population of the ground state hydrogen, $H_2(X^1\Sigma_g^+)$, is important because it can give us information about the production of negative ions, H⁻. Main channel of the creation of H⁻ atoms is usually dissociative attachment of vibrationally excited $H_2$ molecules (Hjartarson *et al* 2010, Bacal and Wada 2015). It was shown that there is a correlation between the evolution of the vibrational temperature of the ground state and the concentration of H⁻ species (Wang *et al* 2008, Kado *et al* 2005). In the following we describe the method used for the determination of the vibrational temperature. It is based on the investigation of the vibrational distribution of $H_2(d^3\Pi_u^-)$ state. This method is sensitive to the first 5 vibrational levels as higher levels are populated by the collisional radiative redistribution via $H_2(C^1\Pi_u^\mp)$ and $H_2(B^1\Sigma_u^+)$ states (Hiskes *et al* 1985). Vibrational distribution of $H_2(d^3\Pi_u^-)$ state can be calculated according to the Franck-Condon principle (Wang *et al* 2008, Fantz and Heger 1998) or by the Gryzinski theory (Xiao *et al* 2004). In general, for the determination of ro-vibrational distributions in hydrogen plasmas at low pressures we can use the simple model of corona equilibrium, based on the assumption that the $H_2(d^3\Pi_u^-)$ state is populated primarily by electron impact from the ground state $H_2(X^1\Sigma_g^+)$ and its deexcitation is caused mainly by radiative transition to the $H_2(a^3\Sigma_g^+)$ excited state. Rotational and vibrational states of $H_2(X^1\Sigma_g^+)$ are assumed to present Boltzmann distributions in this method. At steady state the total production and destruction rates according to the mentioned processes of the ro-vibrationally excited $H_2(d^3\Pi_u^-)$ state are equal and we can write the following kinetic equation

$$\sum_v (K_{Xv}^{dv'} N_{Xv}) = N_{dv'} A_{total}, \qquad (2)$$

where $K_{Xv}^{dv'}$ is electron excitation rate constant from ground state $H_2(X^1\Sigma_g^+, v)$ to $H_2(d^3\Pi_u^-, v')$, $N_{Xv}, N_{dv'}$ are concentrations of the corresponding vibrationally excited states and $A_{total}$ is sum of transition probabilities from the $H_2(d^3\Pi_u^-, v')$ state to $H(a^3\Sigma_g^+, v'')$ for rotational, vibrational and electric part. According to Wang *et al* (2008) and Fantz and Heger (1998), vibrational distribution of the $H_2(d^3\Pi_u^-)$ state can be calculated considering the Franck-Condon approximation

$$N_{dv'} \propto \sum_v q_{Xv}^{dv'} N_{Xv} e^{-\Delta G_{vib}/kT_e}, \qquad (3)$$

where the electron impact excitation rate for the transition $H_2(X^1\Sigma_g^+ \to d^3\Pi_u^-)$ from the ground state with population $N_{Xv}$, is considered to be proportional to the Franck-Condon factors, $q_{Xv}^{dv'}$ (Fantz and Wünderlich 2004), multiplied by an exponential factor which takes into account the threshold of vibrational levels of ground state, $\Delta G_{vib} = G_v - G_0$. $G_v$ and $G_0$ (Fantz and Wünderlich 2004) are the energies of vibrational states *v* and *v* = 0. Finally, $T_e$ denotes the electron temperature. However, Xiao *et al* (2004) called upon the fact that the electron impact excitation does not necessarily obey the Franck-Condon principle, especially for electron temperatures lower than 10 eV. They calculated vibrational distribution according to the following relation

$$N_{dv'} \propto \sum_v k_{Xv}^{dv'} N_{Xv}, \qquad (4)$$

where $k_{Xv}^{dv'}$ is the rate constant for electron impact excitation from the ground state to $H_2(d^3\Pi_u^-, v')$ calculated by the Gryzinski semi classical theory through the determination of the differential excitation cross sections (Bauer and Bartky 1965). If we suppose Boltzmann distribution of vibrational states $H_2(X^1\Sigma_g^+, v)$ the following will be valid

$$N_{Xv} = \frac{exp\left(-\frac{G_v}{kT_{vib}}\right)}{\sum_v exp\left(-\frac{G_v}{kT_{vib}}\right)}, \qquad (5)$$

where $T_{vib}$ denotes the vibrational temperature of the ground state molecule. The figure 9 presents the calculated population ratios of $H_2(d^3\Pi_u^-, v' = 0$ and 1) as function of the electron and vibrational temperature using the Franck-Condon approximation (a) and the Gryzinski theory (b), according to (3) and (4), respectively. The latter presents a maximum which tends to get sharper for decreasing values of $T_e$. As in our conditions the electron temperatures are lower than 10 eV at lower pressure and bi-Maxwellian at higher pressure, the choice of the method is crucial. Due to the low electron temperature measured in our plasma (much below 10 eV), we have chosen to use the method including the Gryzinsky theory, as Xiao *et al* (2004), to calculate the rate constants in order to describe the vibrational distribution of the $H_2(d^3\Pi_u^-, v')$ state.

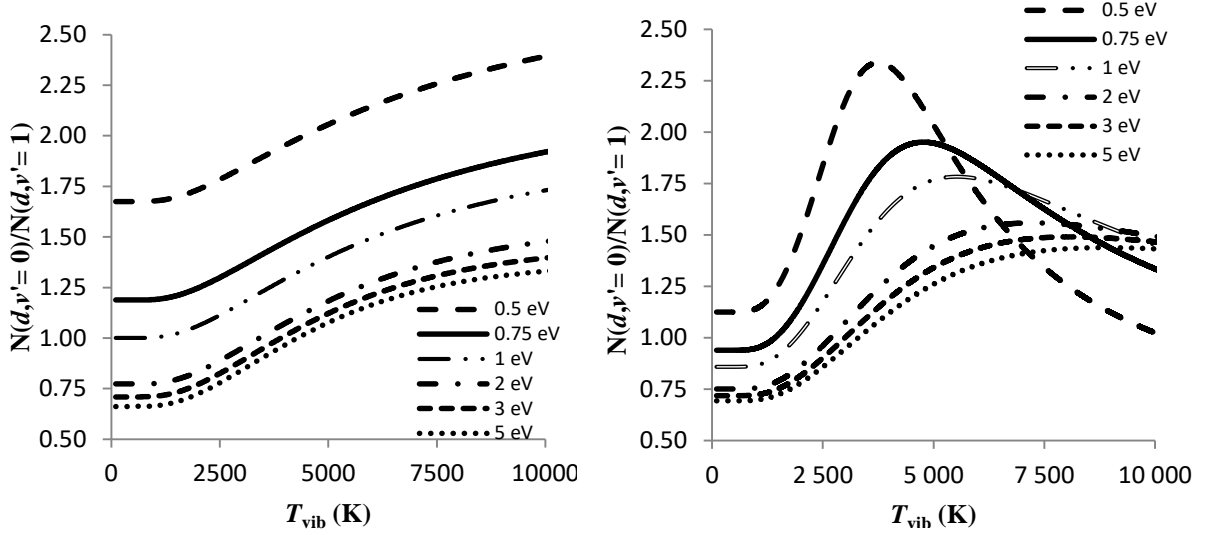

**Figure 9.** Population ratio of the vibrational states $H_2(d^3\Pi_u^-, v' = 0$ and $1)$ calculated according to Franck-Condon principle (a) and Gryzinsky theory (b) as function of the vibrational temperature of the ground state for various electron temperatures. For our calculations the Gryzinsky theory is used. The figures provide the possibility to compare the results with the Franck-Condon principle, for our discharge the incorrect approach.

Concerning the concentration of rotational state $N_{dv'N'}$, we can write

$$N_{dv'N'} = N_{dv'} \frac{(2\Gamma_{N'}+1)(2N'+1)}{\sum_{N'} \exp(-\frac{F_{dv'N'}}{kT_{rot}})} \exp(-\frac{F_{dv'N'}}{kT_{rot}}) \ . \tag{6}$$

Intensity of rotational line $I_{av''N''}^{dv'N'}$ of the Fulcher-α band can be expressed as

$$I_{av''N''}^{dv'N'} = \frac{64\pi^4}{3h\lambda^3} \frac{1}{2N'+1} N_{dv'N'} A_{av''N''}^{dv'N'} \ , \tag{7}$$

where

$$A_{av''N''}^{dv'N'} = S_{aN''}^{dN'} q_{av''}^{dv'} A_a^d \ , \tag{8}$$

where $S_{aN''}^{dN'}$ are the rotational line strength or Hönl–London factor, $q_{av''}^{dv'}$ the Franck-Condon factors of the transition $H_2(d^3\Pi_u^- \to a^3\Sigma_g^+)$ (Fantz and Wünderlich 2004) and $A_a^d$ is the probability of the electronic transition. Intensity can be simplified to

$$I_{av''N''}^{dv'N'} \propto \frac{1}{\lambda^3} q_{av''}^{dv'} N_{dv'N'} \tag{9}$$

then we obtain for the intensity of the Fulcher-α bands lines:

$$I_{av''N''}^{dv'N'} \propto \frac{1}{\lambda^3} \frac{q_{av''}^{dv'}(2\Gamma_{N'}+1)(2N'+1)}{\sum_{N'} \exp(-\frac{F_{dv'N'}}{kT_{rot}})} \exp(-\frac{F_{dv'N'}}{kT_{rot}}) \sum_v \left( K_{Xv}^{dv'} \frac{exp\left(-\frac{G_v}{kT_{vib}}\right)}{\sum_v exp\left(-\frac{G_v}{kT_{vib}}\right)} \right). \tag{10}$$

To calculate vibrational temperature of $H_2(X^1\Sigma_g^+, v)$ using equation (10), we used ratio of rotational lines Q1 from 0-0 and 1-1 bands of $H_2(d^3\Pi_u^-, v')$ and rotational temperature $T_{rot}$ of corresponding vibrational band of the $H_2(d^3\Pi_u^-, v')$ state was used as input parameter.

The figure 10 presents the determined vibrational temperatures of the ground state as function of pressure and delivered power. The error of the method is equal to ~ ± 500 K that comes from uncertainty of rotational temperature and determination intensity of rotational lines. Vibrational temperatures are approximately constant as function of pressure and power for a given mode, but differ markedly in CCP and ICP modes. In the CCP mode the temperature is lower than in the ICP mode by about 1000 K. Overall $T_{vib}$ does not exceed the value of 3500 ± 500 K. Average vibrational temperature of $H_2(X^1\Sigma_g^+, v)$ is 3100 and 2000 ± 500 K in ICP and CCP mode, respectively.

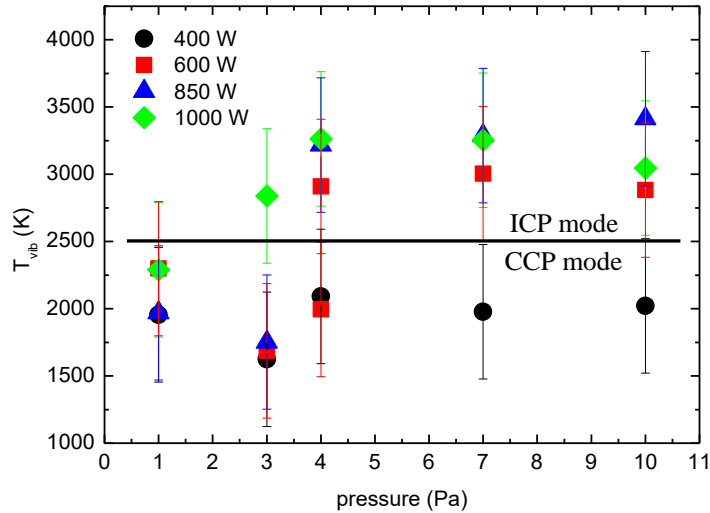

**Figure 10.** Vibrational temperatures of ground state $H_2(X^1\Sigma_g^+)$ as a function of gas pressure for various delivered powers. The samples temperature was held at 300 K (without Pyrex tube). The vibrational temperatures are constant with pressure and power but differ markedly between the two discharge modes.

Within the given uncertainties the value of the vibrational temperature of the ground state in the spatial direction can be considered constant. At 4 Pa, 1000 W and sample temperature 450 K the average value is equal to 3500 and 4100 ± 500 K without and with Pyrex tube, respectively, and at higher pressure (10 Pa, 1000 W and sample temperature 450 K) the averaged vibrational temperature is equal to 3700 and 3300 ± 500 K without and with Pyrex tube, respectively.

Without a complete kinetic modelling of the hydrogen discharge, it is hard to comment the changes of the vibrational temperature with power, distance from the sample or presence of the Pyrex tube. Indeed, many parameters play a role on the population/depopulation of $H_2$ vibrational sates, and still some of them are not well documented. For instance, it is known that vibrational population of $H_2(X^1\Sigma_g^+, v)$ can be influenced by surface recombination of atomic hydrogen that creates vibrationally excited molecules (Cacciatore and Rutigliano 2006, 2009), but the exact amount of energy going into the vibration depends on the surface material itself (Fantz 2002), and is usually not known. Wall deactivation of vibrationnally excited molecules is also a subject of research upon which very few is known (Marinov *et al* 2012, 2014 b, Samuell 2014). Vibrationally excited states are also populated by electron impact and V-V transfers, and depopulated by V-V transfers and V-T transfers with H atoms. Finally, the vibrational population might also be affected by ion- molecules collisions (Xiao *et al* 2005). Many models and studies have been dedicated to the vibrational temperature of $H_2$, taking into account some of the mechanism described above (Mendez *et al* 2006**,** Shakhatov *et al* 2005, Gorse *et al* 1987, Hjartarson *et al* 2010, Xiao *et al* 2005). Let us note that according to Fantz *et al* (2001) electron density has influence on vibrational distribution for electron temperatures higher than 4 eV and electron densities at least $10^{11}$ $cm^{-3}$ . In our study we obtained lower values for ne and Te. Therefore we can suspect an effect of the wall material when the Pyrex tube is introduced and the vibrational temperature is decreased at 4 Pa.

Kalache *et al* (2004) found ~~out~~ that concentration of negative ions is negligible for vibrational temperatures lower than 3000 K. For temperatures higher than 3000 K, concentrations of negative ions increases exponentially. Kimura and Kasugai (2010) supposed creation of negative ions by dissociative attachment from the vibrationally excited hydrogen molecules of $H_2(X^1\Sigma_g^+, v)$ with $v \geq 4$ and their destruction by detachment collision with atomic hydrogen. They characterised concentration of negative ions by electronegativity, α:

$$\alpha = C_{ad} \frac{[H_2(v \geq 4)]}{[H]}, \qquad (11)$$

where $C_{ad}$ is the ratio of rate coefficient for the dissociative attachment to that for the detachment, estimated to be around 10 at maximum (Zorat *et al* 2000, Janev *et al* 1987, Graham 1995). H is the concentration of hydrogen atoms and is obtained from actinometry (see section 3.4.3.). The figure 11 is depicting the calculated values of electronegativity according to equation (11) as function of pressure for two powers. The electronegativity α is below 0.01 in CCP mode and below 0.4 in the case of ICP mode. Thus we can assume that the concentration of negative ions is negligible in CCP mode.

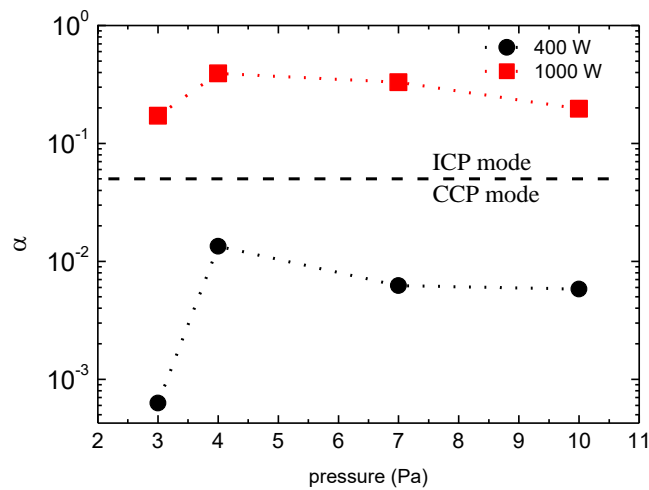

**Figure 11.** Calculated α coefficient for various determined vibrational temperatures of ground state $H_2(X^1\Sigma_g^+)$ as a function of pressure at 400 W (CCP mode) and 1000 W (ICP mode). From the given low values we can assume that the concentration of negative ions is negligible in CCP mode.

*3.4. Estimation of concentration of atomic hydrogen*

We used two actinometric schemes to determine the H atom concentrations by employing either argon or $H_2$ molecules as actinometer. First we discuss the theory behind the two studies (see sections 3.4.1. and 3.4.2.), and then we present the results (see section 3.4.3.).

*3.4.1. Actinometric study with argon admixture.* We used 5 % of argon as actinometer. The basic assumption is that this amount will not change significantly the discharge properties. The following processes were taken into consideration, where *i* and *j* refer to electronic excited states.

- Direct excitation from atomic ground state (IST-Lisbon database):

$$H + e \xrightarrow{k_e^H} H_i + e \quad \text{with } \sigma_e^H(E) \qquad (12)$$

$$Ar + e \xrightarrow{k_e^{Ar}} Ar_i + e \quad \text{with } \sigma_e^{Ar}(E). \qquad (13)$$

- Dissociative excitation (IST-Lisbon database):

$$H_2 + e \xrightarrow{k_d^H} H_i^* + H + e \quad \text{with } \sigma_d^H(E). \qquad (14)$$

- Radiative decay (Physical Meas. Laboratorys database 2014):

$$H_i \xrightarrow{A_{ij}^H} H_j + h\nu_{ij}^H \quad \text{with } A_{ij}^H \qquad (15)$$

$$Ar_i \xrightarrow{A_{ij}^{Ar}} Ar_j + h\nu_{ij}^{Ar} \quad \text{with } A_{ij}^{Ar} \qquad (16)$$

Assuming that any other mechanism has only a negligible contribution to the population/depopulation of excited states, we can write at steady state:

$$\frac{I_H}{I_{Ar}} = \frac{\lambda_{ij}^{Ar} A_{ij}^H (k_e^H [H] + k_d^H [H_2])}{\lambda_{ij}^H A_{ij}^{Ar} k_e^{Ar} [Ar]} \frac{\sum_{k<j} A_{ik}^{Ar}}{\sum_{k<j} A_{ik}^H}. \qquad (17)$$

We supposed that particles with highest concentrations are hydrogen molecules and neutral atomic hydrogen. Concentration of hydrogen molecules can be expressed as

$$[H_2] = \frac{p_{off}}{kT_g} - [H] - [Ar]. \qquad (18)$$

The gas temperature $T_g$ is taken from the measurement, $p_{off}$ is neutral gas pressure in chamber before ignition of discharge given by the capacitance gauge, [Ar] is 0.05*($p_{off}/kT_g$) and rate constants were calculated using EEDF from paragraph 3.1. In actinometry, two transitions having excitation cross sections varying similarly with electron energy are usually employed. Finally, the hydrogen density is given by:

$$[H] = \frac{I_H \lambda_{ij}^H A_{ij}^{Ar} \sum_{k<j} A_{ik}^H k_e^{Ar}[Ar]}{I_{Ar} \lambda_{ij}^{Ar} A_{ij}^H \sum_{k<j} A_{ik}^{Ar} (k_e^H - k_d^H)} - \frac{k_d^H}{k_e^H - k_d^H}\left(\frac{p}{kT_g} - [Ar]\right). \qquad (19)$$

*3.4.2. Actinometric study in pure hydrogen.* The rotational line with $N = 1$ of Q-branch of Fulcher-α band of vibrational transition $v'$-$v''$=2-2 was used as actinometric line. In addition to reactions mentioned before we took into consideration the following ones:

- Direct excitation from molecular ground state:

$$H_2(X, v = 0, N = 1) + e \to H_2(d, v', N = 1) + e \text{ with } \sigma_e^{H_2(d,v,N)}(E) = \sigma_e^{H_2(d,v)}(E) \times a_{N,N'}, \quad (20)$$

where $\sigma_e^{H_2(d,v)}(E)$ was calculated from semi-classical theory of Gryzinski and $a_{N,N'}$ from adiabatic approximation (Xiao *et al* 2004, Farley *et al* 2011).

- Radiative decay (Käning *et al* 1994):

$$H_2(d, v', N = 1) \xrightarrow{A_{d-a}^{H_2(d,v',N=1)}} H_2(a, v'', N) + h\nu_{d-a}^{H_2(d,v',N=1)} \text{ with } A_{d-a}^{H_2(d,v,N=1)}. \quad (21)$$

Intensity of hydrogen excited state and rotational line $N = 1$ of Q branch of Fulcher-α band is given:

$$I_{Q1} = const \cdot h\nu_{H_2(d,v,1)} A_{d-a}^{H_2(d,v',1)} [H_2(d, v', 1)], \quad (22)$$

where

$$[H_2(d, v', 1)] = k_e^{H_2(d,v',1)} [H_2(X, v = 0, N = 1)] n_e \frac{1}{\sum_{k<j} A_{ik}^{H_2(d,v',1)}}. \quad (23)$$

Ratio of atomic hydrogen line and rotational line of Fulcher-α band will be equal to

$$\frac{I_H}{I_{Q1}} = \frac{\lambda^{Q1} A_{ij}^H \ (k_e^H \ [H] + k_d^H \ [H_2])}{\lambda_{ij}^H \ A_{d-a}^{H_2(d,v',1)} k_e^{H_2(d,v',1)} [H_2(X,v=0,N=1)]} \frac{\sum_{k<j} A_{ik}^{H_2(d,v',1)}}{\sum_{k<j} A_{ik}^H}, \quad (24)$$

where we can write that

$$[H_2] = \frac{p}{kT_g} - [H] \quad (25)$$

and

$$\eta(T_g) = \frac{H_2(X,v=0,N=1)}{H_2} = \frac{9}{4} \frac{1}{\sum_{N=1,3,5,...}(2N+1)\exp\left(\frac{-(E_N-E_1)}{kT_g}\right)}. \quad (26)$$

Where $T_g$ is assumed to be equal to $T_{rot}$. The formula (26) according to Lavrov *et al* (2003) describes the influence of rotational excitation on population of the rotational level $N=1$. $E_N$ is energy of the rotational level with the rotational quantum number $N$ in the $v = 0$ ground vibrational state of molecular hydrogen. Finally, it comes:

$$[H] = \frac{\frac{I_H \lambda_{ij}^H A_{d-a}^{H_2(d,v',1)} k_e^{H_2(d,v',1)} \sum_{k<j} A_{ik}^H}{I_{Q1} \lambda^{Q1} A_{ij}^H \sum_{k<j} A_{ik}^{H_2(d,v',1)}} k_e^{H_2(d,v',1)} \eta(T_g) \frac{p}{kT_g} - k_d^H \frac{p}{kT_g}}{\frac{I_H \lambda_{ij}^H A_{d-a}^{H_2(d,v',1)} k_e^{H_2(d,v',1)} \sum_{k<j} A_{ik}^H}{I_{Q1} \lambda^{Q1} A_{ij}^H \sum_{k<j} A_{ik}^{H_2(d,v',1)}} k_e^{H_2(d,v',1)} \eta(T_g) + k_e^H - k_d^H}. \quad (27)$$

*3.4.3. Comparison of results from the two actinometric studies.* The figure 12 (a) and (b) presents the evolution of hydrogen atoms concentration with pressure for 400 and 1000 W calculated using argon or $H_2$ as actinometers. In the case of argon, the Ar transition from $2p_1$ to $1s_2$ at 750.4 nm has been used and two different H transitions have been employed: H-alpha line at 656 nm (figure 12 (a)) and H-beta line at 486 nm (figure 12 (b)). For hydrogen, Q branch of Fulcher-α band has been used.

First we can note that the two hydrogen transitions are giving different results. Actinometry using H-alpha line results higher concentrations of atomic hydrogen than using H-beta line. Similar results were observed by Skoro *et al* (2013). They observed better agreement between atomic hydrogen concentrations determined from actinometry and from catalytic probe measurements when using H-beta line rather than H-alpha. Therefore we have considered the H-beta line for further calculations. We can see also, that Fulcher-$\alpha$ and argon actinometry (H-beta) are in good agreement (figure 12 (b)) which is not the case for the results obtained with H-alpha line (figure 12 (a)), confirming that the use of H-alpha should be avoided here. The presented graph shows the evolution of H atom concentrations in the CCP and ICP modes with pressure between 2 and 10 Pa. In general, the H atom concentration is increasing with the pressure. Surprisingly, we have obtained higher H concentration in the CCP mode than in ICP mode, with densities ranging from around $1.5 \times 10^{13}$ to $7.5 \times 10^{13}$ cm$^{-3}$ in CCP and from $1 \times 10^{13}$ to $4 \times 10^{13}$ cm$^{-3}$ in the ICP mode with increasing pressure from 3 to 10 Pa. This point is discussed later. Relative to previous works, comparable values of hydrogen concentration were found by Mendez *et al* (2006) in a DC discharge with values between $10^{13}$-$10^{14} \times$cm$^{-3}$ in pressure range 1-10 Pa at electron densities between $1$-$3 \times 10^{10}$ cm$^{-3}$ and decreasing electron temperature from 8 to 2 eV. Marques *et al* (2007) measured a value of $7 \times 10^{13}$ cm$^{-3}$ but at higher pressure, 26 Pa, with electron density up to $4 \times 10^9$ cm$^{-3}$ and electron temperature of 3 eV in a CCP RF discharge at 13.56 MHz. At lower pressure, 1 Pa, in a helicon type reactor Skoro *et al* (2013) determined concentration of hydrogen atoms between $5$-$9 \times 10^{12}$ cm$^{-3}$ at powers ranging between 800 and 1000 W.

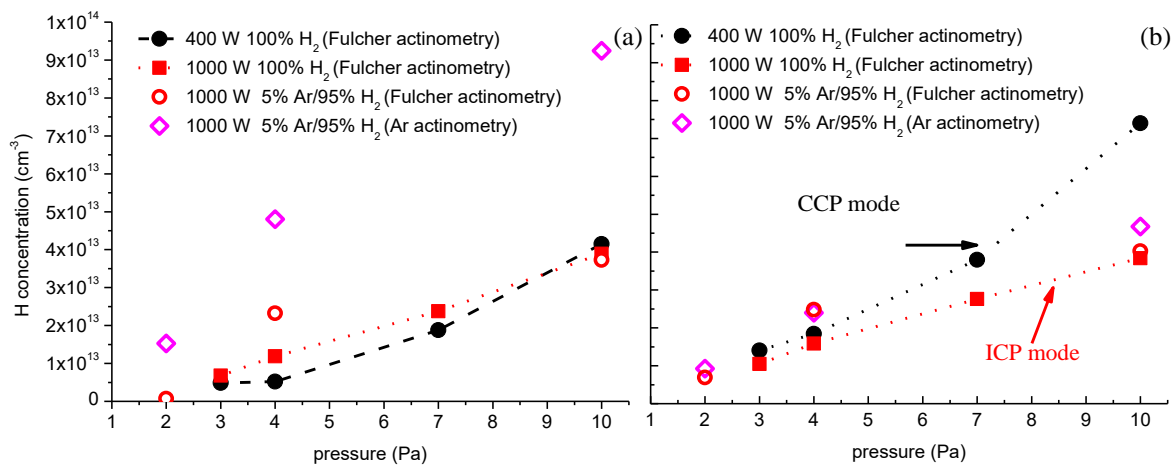

**Figure 12.** Evolution of atomic hydrogen concentrations with pressure at 400 and 1000 W while the sample holder temperature was held at 300K (without sample). On graphs (a) and (b) we are presenting the results of actinometric technique using the H-alpha and H-beta lines, respectively. The Fulcher-$\alpha$ and argon actinometry with H-beta line are in good agreement, the use of H-alpha should be avoided. The resulted H atom concentration is increasing with the pressure.

The figure 13 is depicting the evolution of the dissociation ratio as function of pressure at 400 and 1000 W representing the two discharge modes. Dissociation ratio increases from 2.7 % at 3 Pa to 4.1

% at 10 Pa in ICP mode and from 2.3 % at 3 Pa to 3.5 % at 10 Pa in CCP mode. We can observe higher dissociation ratio in the ICP mode than CCP mode. At first sight this seems to be in disagreement with the evolution of H atom concentrations presented on figure 12 (b), where we obtain lower H atom concentrations in ICP mode compared to the CCP mode. In fact, the lower concentrations in the ICP mode are caused by increasing gas temperature which induces a decrease of hydrogen molecule density (and therefore of hydrogen dissociation). Still, it is surprising that dissociation degree in ICP and CCP mode are so close while electron density are differing by one order of magnitude. This might be due to the increased ion flux to the wall that leads to a higher atomic loss at the wall as observed or suggested in different works (Samuell and Corr 2015, Cartry *et al* 1999, Jolly and Booth 2005). Regarding the literature, the obtained dissociation is comparable with results of other works. In the work of Hjartarson *et al* (2010) the dissociation was 25 % at 0.53 Pa, 20 % at 1 Pa and 5 % at 13Pa in pure $H_2$ discharge. Dissociation ratio decreased from 17 % to 7 % in pressure range 0.8 to 20 Pa according to the article of Mendez *et al* (2006) combining experimental and theoretical study on a hollow cathode DC discharge.

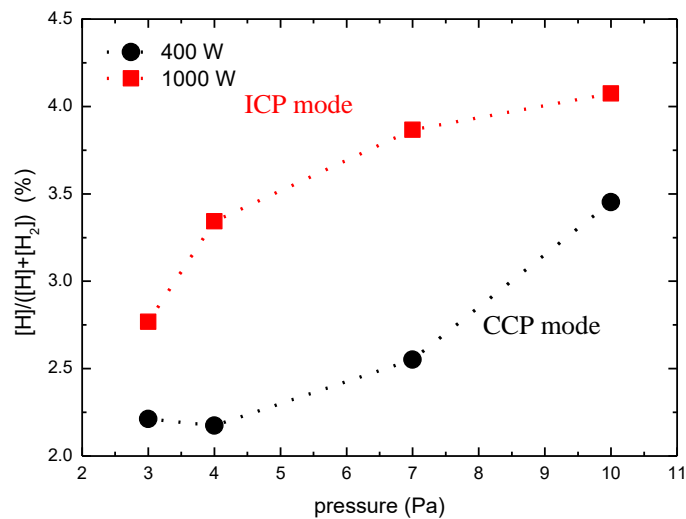

**Figure 13.** Dissociation ratio as function of pressure at 400 and 1000 W. The dissociation ratio is higher in the ICP mode than CCP mode.

We also measured the temporal evolution of the H atom concentrations in the pulsed mode. We observed an initial fast increase (shorter than 5 ms), and then the H density are reaching more or less the value reached in non-pulsed mode. At the end of the discharge pulse, the H density reaches $1.5 \times 10^{13}$ and $4.5 \times 10^{13}$ cm$^{-3}$ at 4 and 10 Pa, respectively.

The figure 14 presents the evolution of the H atom concentrations as function of distance from the sample holder at 4 and 10 Pa with delivered power of 1000 W. Argon actinometry with H-beta was used here. We observe higher density with the tube than without. This is due both to the higher power density (discharge volume is smaller with the tube) and to the use of Pyrex as wall material (rather than stainless steel) allowing to reduce atomic loss. Indeed, surface loss on glass materials is usually much lower than on metals, ~ one order of difference (e.g. for glass: Cartry *et al* 2000, Macko *et al* 2004, Rousseau *et al* 2001; e.g. for metals: Sode *et al* 2014). The concentration at 10 Pa is higher than at 4 Pa by a factor between 2.5 a 5. For both pressures there is a slight rise in atomic concentration with increasing distance from the sample holder. It is correlated to the loss of atoms at the sample surface. Loss coefficient can theoretically be determined from the slope. However, the determination is quite hard in presence of temperature gradients (Booth *et al* 2005) as it is in the case in the present study.

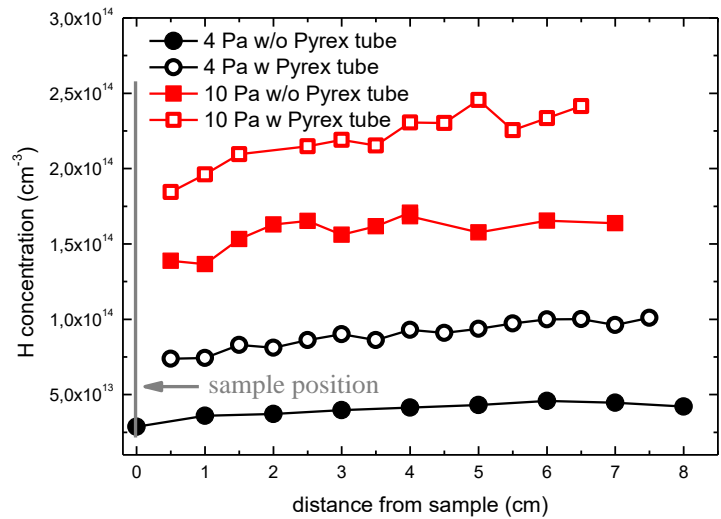

**Figure 14.** Evolution of atomic hydrogen concentration at 4 Pa (black colour) and at 10 Pa (red colour) with Pyrex tube (empty circle) and without Pyrex tube (full circle) at power of 1000 W and temperature of sample holder 450 K. The gas mixture was $H_2$ / 5 % Ar. H-beta line has been used. The H atom concentration is higher with the tube due to higher power density and the choice of wall material, as Pyrex is reducing atomic loss.

## 4. Conclusion

This study presents a preliminary characterisation of a hydrogen plasma source, proposed for the determination of the reflection coefficient for fusion related studies. Within the framework of this project, a hydrogen RF discharge was studied at frequency of 13.56 MHz, at pressures between 4 and 10 Pa, and for power in range of 400 - 1000 W. The characterization of the plasma was realized by means of optical emission spectroscopy in both CCP and ICP discharge modes. Moreover Langmuir probe measurements were carried out in order to determine the electron energy distribution function and electron density. Some measurements were realized in admixtures of nitrogen up to 50 % and actinometry measurements in 5 % argon. Additionally, for some cases of the studied discharge parameters the analysis of a pulsed discharge with duty ratio 50:50 has been included. The experimental set-up allowed the investigation of the plasma in the presence of a sample or in the absence of it. For some measurements a Pyrex tube was used which delimited the volume of the discharge, allowing for a better definition of wall conditions for surface loss measurements. Note, that the placement of the sample ($SiO_2$) within the experimental set-up allowed the interaction with the plasma, while the sample holder (made of molybdenum) could be heated to a desired temperature.

First of all, the electron densities were observed to be increasing with both pressure and power. The transition to the ICP mode was characterized by a sudden jump in the electron densities above ~ $10^{10}$ cm$^{-3}$, observed for both pure hydrogen and in a $H_2$-$N_2$ discharge. The energy distribution function of electrons was found to be a Maxwellian one at low pressure, but for increasing pressure it gradually changed to a bi-Maxwellian one at around 4 Pa. There was no power dependence observed.

The gas temperature in the CCP discharge mode was constant with pressure and power with an average value of 450 ± 50 K. In the ICP mode the gas temperature was increasing with increasing pressure and power, in accord with the literature. These values were obtained from the Fulcher-α system of hydrogen, but also confirmed by $H_2$-$N_2$ mixtures by the second positive system of nitrogen. For the case of a pulsed discharge, the temperature was increasing during the pulse and reached a stationary value after ~1.5 ms, which corresponded to the value of the continuous discharge.

The vibrational temperature of the ground state $H_2(X^1\Sigma_g^+, v)$ molecule was determined using the Gryzinski theory. The corresponding functions were found to be approximately constant with respect to both pressure and power for a given mode, but differ significantly between CCP and ICP modes. Average vibrational temperature of $H_2(X^1\Sigma_g^+, v)$ was 3100 and 2000 ± 500 K in ICP and CCP mode respectively. The vibrational temperature was also studied as a function of distance from the sample.

Based on calculations of the electronegativity, found to be 0.01 in CCP mode and below 0.4 in ICP mode, and also referring to previous studies, the concentration of negative ions is assumed to be negligible in CCP mode.

Atomic hydrogen concentration was determined by actinometry using 5 % of Ar as a probing gas, or by using the Q1 rotational line of Fulcher-α system as actinometer. Concentration of hydrogen density was increasing with pressure in both modes (from around 1.5×10$^{13}$ to 7.5×10$^{13}$ cm$^{-3}$ in CCP and from 1×10$^{13}$ to 4×10$^{13}$ cm$^{-3}$ in the ICP mode with increasing pressure from 3 to 10 Pa), but with a dissociation degree slightly higher (by a factor of 2) in the ICP mode. In the case of a pulsed discharge, an initial fast increase of atomic concentration was observed. Then, at around 5 ms, the value reached the value of the continuous mode.

The presented study will be a starting point for further calculations of reflection coefficient versus experimental conditions. This kind of study requires a good knowledge and good characterization of the plasma for two main reasons. The first one concerns the measurement of the loss coefficient in itself that can be badly estimated if effects such as gas temperature and neutral density variations are not well managed (Cartry *et al* 2006, Lamara *et al* 2006). Second, a key point in surface loss studies is the identification of the parameters playing a role on the surface loss coefficient. The external control "knobs" of the plasma (pressure, power…) affect several plasma parameters and several species concentration at the same time. Unraveling the influence of all plasma parameters on the surface loss coefficient thus requires a complete characterization of the plasma versus the experimental conditions. Using the present experimental device, the ion flux can be varied in a controlled manner, the gas temperature is known and most of the plasma parameters are estimated. This is a pre-requisite for H atomic loss coefficient studies.

**Acknowledgments.** The authors thank for financial support from the Scientific Grant Agency of the Slovak Republic (VEGA) under the contract No. 1/0925/14 and No. 1/0914/14 and also from the Slovak Research and Development Agency under the project No. DO7RP-0021-12. This work was carried out within the framework of the French Research Federation for Fusion Studies (FR-FCM) and of the EUROfusion Consortium. It has received funding from the European Union's Horizon 2020 research and innovation program under grant agreement No. 633053. The views and opinions expressed herein do not necessarily reflect those of the European Commission. Work performed under EUROfusion WP PFC.